\documentclass[a4paper,11pt]{article}
\pdfoutput=1 

\usepackage{jcappub} 

\usepackage[T1]{fontenc} 

\title{Relativistic weak lensing from a fully non-linear cosmological density field}

\author[a,b,1]{D.B.Thomas,\note{Corresponding author.}}
\author[b]{M. Bruni,}
\author[b]{D. Wands}

\affiliation[a]{Department of Physics, University of Cyprus, Aglantzia, Nicosia, 2109}
\affiliation[b]{Institute of Cosmology and Gravitation, University of Portsmouth, Dennis Sciama Building, Burnaby Road, Portsmouth, PO1 3FX, UK}

\emailAdd{thomas.daniel@ucy.ac.cy}
\emailAdd{marco.bruni@port.ac.uk}
\emailAdd{david.wands@port.ac.uk}

\abstract{In this paper we examine cosmological weak lensing on non-linear scales and show that there are Newtonian and relativistic contributions and that the latter
can also be extracted from standard Newtonian simulations. We use the post-Friedmann formalism, a post-Newtonian type framework for cosmology,
to derive the full weak-lensing deflection angle valid on non-linear scales for any metric theory of gravity. We show that the only contributing
term that is quadratic in the first order deflection is the expected Born correction and lens-lens coupling term. We use this deflection angle to
analyse the vector and tensor contributions to the E- and B- mode cosmic shear power spectra. In our approach, once the gravitational theory has
been specified, the metric components are related to the matter content in a well-defined manner. Specifying General Relativity, we write down a
complete set of equations for a GR$+\Lambda$CDM universe for computing all of the possible lensing terms from Newtonian N-body simulations. We
illustrate this with the vector potential and show that, in a GR$+\Lambda$CDM universe, its contribution to the E-mode is negligible with respect
to that of the conventional Newtonian scalar potential, even on non-linear scales. Thus, under the standard assumption that Newtonian N-body
simulations give a good approximation of the matter dynamics, we show that the standard ray tracing approach gives a good description for a
$\Lambda$CDM cosmology.}

\begin{document}

\maketitle
\flushbottom

\section{Introduction}
It has been understood since the dawn of General Relativity that one of the consequences of a gravitational theory incorporating the
equivalence principle is the bending of light by gravitating masses \citep{LorEinMin52}. An important consequence of this light bending
is that galaxy images are distorted by gravitational influences along the line of sight. A statistical analysis of the shapes of galaxy
images allows this weak gravitational lensing effect, known as cosmic shear, to be observed. Since the first measurements of cosmic
shear \citep{Wittman:2000tc,Kaiser:2000if,Bacon:2000sy,vanWaerbeke:2000rm}, increasingly precise measurements have been made of the cosmic
shear power spectra, leading up the the latest CHTLens results \citep{Kilbinger:2012qz}. Now, weak lensing is becoming an important tool
for cosmology, with claims that it can place constraints on the equation of state of dark energy \citep{Albrecht:2006um} and modified
gravity models \citep{euclid}. Current and future surveys include the Dark Energy Survey \citep{Abbott:2005bi}, LSST \citep{Ivezic:2008fe},
SKA \citep{ska} and the ESA satellite mission Euclid \citep{Refregier:2010ss}. These missions will yield a wealth of precise data,
including tomographic weak lensing power spectra. However, these power spectra will largely come from non-linear scales in the universe,
i.e. scales smaller than around 8-10 h$^{-1}$ Mpc, where the matter density contrast has gone non-linear.

Broadly speaking, cosmologists study linear and non-linear scales in different ways. On larger, linear scales, fully relativistic
perturbative schemes are used where the matter and metric perturbations in the universe are assumed to be small. On smaller scales
in the universe, gravity is assumed to be Newtonian and N-body simulations are run to examine the evolution of the density field.
Weak-lensing calculations have been carried out both in the Newtonian regime \citep{Bartelmann:1999yn} and with GR perturbation theory
(see e.g. \citep{dodelson:2003,Dodelson:2005zj}). However, there are few studies that both apply on smaller, non-linear scales in
cosmology and are fully relativistic, thus justifying the standard ray tracing approach to extracting lensing observables from N-body simulations.

The post-Friedmann formalism, introduced in \citep{postf,thesis} and applied in \citep{Bruni:2013mua}, is a non-linear approximation scheme
designed to study relativistic structure formation in the universe on all scales, including scales where the density contrast is large. It
is based upon a post-Newtonian type expansion in inverse powers of the speed of light $c$, building the relativistic corrections on top of
the Newtonian equations. The resulting equations are a non-linear approximation to the full Einstein equations that are valid on all scales.
At leading order, the post-Friedmann formalism reproduces Newtonian cosmology \citep{postf,thesis}.

In this paper, we use the post-Friedmann metric to compute the complete weak-lensing deflection angle up to order $c^{-4}$, including terms
that are quadratic in the first order deflection, the first post-Newtonian corrections to the two scalar gravitational potentials and the 
effects of the vector and tensor contributions to the metric. This is the first time that the complete weak-lensing deflection angle on
non-linear scales has been computed in a fully relativistic manner, including scalar, vector and tensor contributions to the metric. We
then examine how the different terms in the deflection angle contribute to the E- and B-mode power spectra. The calculation of the deflection
angle here is purely geometrical, i.e. does not assume the Einstein equations, thus it should be valid for any metric theory of gravity.

The strength of this approach to weak lensing is that the metric components that contribute to the lensing can be consistently related to the
matter inhomogeneities. We illustrate this with a set of equations demonstrating how each of the quantities that contributes to the deflection
angle can be computed from N-body simulations for a $\Lambda$CDM cosmology and demonstrate that the contribution to the E-mode from the vector
potential is negligible for such a cosmology. The equation for the deflection angle, equation (\ref{bend_final}), coupled with equations
(\ref{complete_matter_eqns1})-(\ref{complete_matter_eqns8}) demonstrating how to relate the metric perturbations to the matter in a $\Lambda$CDM,
comprise the main results of this paper.

This paper is laid out as follows. In section \ref{sec_postf} we will briefly introduce the relevant parts of the post-Friedmann formalism,
which will then be used to calculate the weak-lensing deflection angle in section \ref{sec_bend}. In section \ref{sec_power}, we will examine
how the vector and tensor parts of the deflection angle contribute to the E- and B-mode power spectra of cosmic shear. Section \ref{sec_matter}
will examine how to relate the matter inhomogeneities in the universe to the metric potentials, including the example of the post-Friedmann vector
potential. We conclude in section \ref{sec_conc}.

\section{Post-Friedmann Formalism}
\label{sec_postf}
The post-Friedmann formalism is a new method that has been developed in \citep{postf} for the purpose of studying relativistic structure
formation on all scales. It consists of performing an expansion in inverse powers of the speed of light $c$ in an analogous fashion to
post-Newtonian theory in the Solar System \citep{weinbook}. However, the expansion is performed differently in a cosmological setting,
see \citep{postf} for the full details. Note that one of the differences compared to the Solar System treatment is that the two scalar
potentials are both included at leading order and are not assumed to be equal a priori. The post-Friedmann approach considers a dust
(pressureless matter) cosmology with a cosmological constant and the perturbed FLRW metric, in Poisson gauge, is explicitly written in
terms ordered in powers of $c^{-1}$. Including terms up to $c^{-5}$, the metric thus looks as follows,
\begin{eqnarray}
 g_{00}&=&-\left[1-\frac{2U_N}{c^2}+\frac{1}{c^4}\left(2U^2_N-4U_P \right)\right]\nonumber\\
g_{0i}&=&-\frac{aB^N_i}{c^3}-\frac{aB^P_i}{c^5}\label{metric1}\\
g_{ij}&=&a^2\left[\left(1+\frac{2V_N}{c^2}+\frac{1}{c^4}\left(2V^2_N+4V_P \right) \right)\delta_{ij} +\frac{h_{ij}}{c^4}\right]\nonumber \rm{.}
\end{eqnarray}
Since this metric is in the Poisson gauge, the three-vectors $B^N_i$ and $B^P_i$ are divergenceless, $B^N_{i,i}=0$ and $B^P_{i,i}=0$. In addition,
$h_{ij}$ is transverse and tracefree, $h^i_i=h_{ij}^{,i}=0$. Note that at this order, $h_{ij}$ is not dynamical, so it does not represent gravitational
waves. Rather, $h_{ij}$ represents the transverse traceless part of the metric necessarily arising from non-linear effects. The $g_{00}$ and $g_{ij}$
scalar potentials have been split into their leading order, Newtonian, components ($U_N$, $V_N$) and their post-Friedmann components ($U_P$, $V_P$).
Similarly, the vector potential has been split up into $B^N_i$ and $B^P_i$. From a post-Friedmann viewpoint, there are two different levels of
perturbations in the theory, corresponding to terms of order $c^{-2}$ and $c^{-3}$, or of order $c^{-4}$ and $c^{-5}$ respectively. Inserting the
above metric in Einstein's equations and keeping only the leading order terms shows that $U_N=V_N$ and leads to the standard Newtonian cosmological
equations, with a bonus, the vector potential $B^N_i$ is necessarily non-zero \citep{thesis,postf,Bruni:2013mua}.

Defining ``resummed" variables, such as $\Phi=2U_N+c^{-2}\left(2U^2_N-4U_P \right)$, then calculating the Einstein equations and linearising them,
reproduces linear GR perturbation theory in Poisson gauge. Thus, this approach is capable of describing structure formation on the largest scales. 
The energy momentum tensor is constructed from the dust four-velocity and is also expanded in powers of $c^{-1}$, allowing the Einstein equations
to be derived at different orders in the expansion. Crucially, whilst performing calculations using this $c^{-1}$ expansion, no assumptions are made
regarding the amplitude of perturbations, notably the density contrast, so this expansion and the corresponding equations are valid even on scales
where the density contrast is greater than unity. This will allow us to use the metric and Einstein equations from this formalism to calculate
weak-lensing spectra on fully non-linear scales. The leading order gravitational equations in the $c^{-1}$ expansion and some of the higer order
gravitational equations are introduced in section \ref{sec_matter} for relating the metric potentials to N-body simulations. The derivation of
these equations in a $\Lambda$CDM cosmology is in \citep{postf}.

\section{The Deflection Angle}
\label{sec_bend}
In this section we will calculate the full weak-lensing deflection angle in the post-Friedmann formalism, up to order $c^{-4}$. By truncating at
this order, we include the leading order contributions from each of the possible post-Newtonian corrections, $U_P$, $V_P$, $B^N_i$ and $h_{ij}$.
We will follow the method outlined in \citep{Dodelson:2005zj}, based on the second order geodesic equation from \citep{Pyne:1995bs}, but we will
adapt the method for an expansion in powers of $c^{-1}$. For this work, when examining the Christoffel symbols, we found that we can treat
post-Friedmann $c^{0}$ and $c^{-1}$ quantities analogously to zeroth order in perturbation theory, $c^{-2}$ quantities analogously to first
order in perturbation theory and $c^{-3}$ or $c^{-4}$ quantities analogously to second order in perturbation theory. This will ensure that we
include all of the terms that are relevant for our calculation\footnote{Note that we have categorised the terms this way for this calculation
to simplify it and make it more easily comparable to perturbation theory. If extending this calculation to higher orders, it would be necessary
to treat post-Friedmann $c^{-2}$ and $c^{-3}$ quantities analogously to first order in perturbation theory, $c^{-4}$ and $c^{-5}$ quantities as
second order in perturbation theory, and so on.}. A second order treatment of the geodesic equation is sufficient for a post-Friedmann calculation
up to order $c^{-4}$, since every metric potential is at least of order $c^{-2}$, so any third order terms will be at least of order $c^{-6}$.

We will then compare our result to related calculations in the literature. In practice, what we will see is that the deflection angle from the
post-Friedmann expansion up to order $c^{-4}$ is similar to the second order calculation from perturbation theory. The key differences are: The
leading order term is composed of the Newtonian scalar potential arising from the fully non-linear density field, as opposed to the linear potential
from perturbation theory; there is a post-Friedmann correction to the scalar potential and, a priori, the vector and tensor terms cannot be neglected;
and the equations relating the metric perturbations in the deflection angle formula to the matter inhomogeneities on non-linear scales will also be
different compared to perturbation theory.

\subsection{Post-Friedmann deflection angle calculation}
The starting points for the deflection angle calculation are the metric above \ref{metric1}, truncated at order $c^{-4}$,
\begin{eqnarray}
 g_{00}&=&-1+\frac{2U_N}{c^2}-\frac{1}{c^4}\left(2U^2_N-4U_P \right)\nonumber\\
g_{0i}&=&-\frac{aB_i}{c^3}\label{metric2}\\
g_{ij}&=&a^2\left[\left(1+\frac{2V_N}{c^2}+\frac{1}{c^4}\left(2V^2_N+4V_P \right) \right)\delta_{ij} +\frac{h_{ij}}{c^4}\right]\rm{,}
\end{eqnarray}
and the spatial parts of the geodesic equation
\begin{equation}
 \frac{d^2 x^i}{d\lambda^2}=-\Gamma^{i}_{\mu \nu}\frac{dx^{\mu}}{d\lambda}\frac{dx^{\nu}}{d\lambda}\rm{.}
\end{equation}
The Christoffel symbols for this metric, ordered as described above, are in appendix \ref{app_quant}. We will be solving the geodesic equation
following \citep{Dodelson:2005zj}. Integrating the geodesic equation shows the transverse deflection of the light ray to be given by
\begin{eqnarray}
 x^{(1)I}(\vec{\theta})&=&\int^{\chi_s}_{0}d\chi\left(\chi_s-\chi \right)f^{(1)I}(\chi,\vec{\theta})\nonumber \\
 x^{(2)I}(\vec{\theta})&=&\int^{\chi_s}_{0}d\chi\left(\chi_s-\chi \right)f^{(2)I}(\chi,\vec{\theta})\rm{,}
\end{eqnarray}
where the number in brackets represents the order, in a perturbative sense, of the deflection. The co-ordinates are such that $\chi$ is the comoving
distance and is the $z$ co-ordinate, $\vec{\theta}$ is the transverse displacement from the $z$ axis and the $x$ and $y$ co-ordinates can be written
as $\chi \theta_1$ and $\chi \theta_2$ respectively. The upper limit of the integral, $\chi_s$, is the comoving distance that the photon originated from.
From here onwards, a lower case latin letter will be used to denote 3 dimensional spatial indices (i=1,2,3) and an upper case latin letter will be used
to denote indices that are one of the transverse spatial directions only (I=1,2). The terms in the integral are given by
\begin{eqnarray}
 f^{(1)I}(\chi,\vec{\theta})&&=-\Gamma^{(1)I}_{\alpha\beta}p^{(0)\alpha}p^{(0)\beta}\nonumber \\
f^{(2)I}(\chi,\vec{\theta})&&=-\Gamma^{(0)I}_{\alpha\beta}p^{(1)\alpha}p^{(1)\beta}-2\Gamma^{(1)I}_{\alpha \beta}p^{(0)\alpha}p^{(1)\beta}\nonumber\\
&&-2\partial_{\sigma}\Gamma^{(0)I}_{\alpha\beta}x^{(1)\sigma}p^{(0)\alpha}p^{(1)\beta}-\partial_{\sigma}
\Gamma^{(1)I}_{\alpha \beta}x^{(1)\sigma}p^{(0)\alpha}p^{(0)\beta}\nonumber\\
&&-\frac{1}{2}\partial_{\sigma}\partial_{\tau}\Gamma^{(0)I}_{\alpha\beta}x^{(1)\sigma}x^{(1)\tau}p^{(0)\alpha}p^{(0)\beta}
-\Gamma^{(2)I}_{\alpha\beta}p^{(0)\alpha}p^{(0)\beta}\label{eq_f2}\rm{,}
\end{eqnarray}
where $p^{(i)\alpha}$ is the photon direction vector and at zeroth order is normalised to $p^{(0)\alpha}=\left(-a,0,0,1 \right)$.
\footnote{Note that the $p^{(0)0}$ component here is different to \citep{Dodelson:2005zj} due to the use of cosmic time in our metric (\ref{metric2}).}
This is because we are working in the small angle approximation and means that we can ignore terms involving $p^{(0)I}$.  The first order
photon direction vector ($p^{(1)\alpha}$), deflection ($x^{(1)\alpha}$) and geodesic terms ($f^{(1)\alpha}$) are all in appendix \ref{app_quant},
these will be needed for calculating the six terms contributing to $f^{(2)I}$. We will examine the six terms individually, discarding any terms
of order $c^{-5}$ or above, before combining them into the final deflection angle.\\

Noting that each $p^{(1)\alpha}$ or $x^{(1)\alpha}$ term includes a factor of $c^{-2}$ and that all zeroth order Christoffel symbols include a factor
of $c^{-1}$ due to the time derivative, a quick examination shows that the first, third and fifth terms are all at least of order $c^{-5}$, so can be neglected.

%
%

These terms serve to illustrate how, whilst the post-Newtonian expansion up to order $c^{-4}$ is similar to second order perturbation theory,
some of the terms don't survive. All three of these terms are at least of order $c^{-5}$ because the zeroth order (background) universe is
spatially homogeneous, meaning that the only derivative of the zeroth order universe that survives is the time derivative. This is considered
to be higher order than a spatial derivative in the post-Friedmann formalism, so has a factor $c^{-1}$ associated with it.

The remaining second, fourth and sixth terms in equation (\ref{eq_f2}) will all contribute to the deflection angle at $c^{-4}$. In all three of
these terms, we will make use of $p^{(0)I}=0$.\\

For the second term, $-2\Gamma^{(1)I}_{\alpha \beta}p^{(0)\alpha}p^{(1)\beta}$, the $\alpha=0,\beta=j$ and $\alpha=3,\beta=0$ terms are both
zero for our purposes: For the first of these, the first order Christoffel symbol $\Gamma^{(1)i}_{0j}$ is of order $c^{-3}$ and the $p^{(1)j}$ is
of order $c^{-2}$ making the term of order $c^{-5}$. For the second of these, the 1st order Christofel symbol $\Gamma^{(1)I}_{03}=0$. Then the
remainder of this second term is
\begin{eqnarray}
-2\Gamma^{(1)I}_{\alpha \beta}p^{(0)\alpha}p^{(1)\beta}&&=\frac{4}{c^4}U^{,I}_{N}\int^\chi_0d\chi'U_{N,3}
-\frac{1}{c^4}V_{N,3}\int^{\chi}_0d\chi'\left(U^{,I}_{N}+V^{,I}_{N}\right)\nonumber\\
&&+\frac{2}{c^4}V^{,i}_N\int^{\chi}_0d\chi'\left(U_{N,3}-V_{N,3}\right)\rm{.}
\end{eqnarray}

The fourth term contains the first order Christoffel symbol, which is non-zero if the lower indices are both spatial or both temporal.
In addition, the $\partial_{0}$ term can also be ignored because of the extra factor of $c^{-1}$, which makes the term of order $c^{-5}$.
\begin{eqnarray}
-\partial_{\sigma}\Gamma^{(1)I}_{\alpha \beta}x^{(1)\sigma}p^{(0)\alpha}p^{(0)\beta}
&&=\frac{\left(U_{N}+V_{N} \right)^{,I}_{,J}}{c^4}\int^{\chi}_{0}\left(\chi-\chi' \right)\left[U^{,J}_N+V^{,J}_N\right]d\chi'\nonumber\\
&&+\frac{\left(U_{N}+V_{N} \right)^{,I}_{,3}}{c^4}\int^{\chi}_{0}\left(\chi-\chi' \right)\left[U^{,3}_N-V^{,3}_N\right]d\chi'
\end{eqnarray}

The 6th term contains the ``true'' 2nd order quantities due to the second order Christoffel symbol, including the vector and tensor contributions.
 \begin{eqnarray}
  -\Gamma^{(2)I}_{\alpha\beta}p^{(0)\alpha}p^{(0)\beta}&&=\frac{B^{,I}_3-B^I_{,3}}{c^3}+\frac{a}{c^4}\dot{B}^I
  +\frac{1}{2c^4}h^{,I}_{33}-\frac{1}{c^4}h^I_{3,3}\nonumber\\
&&+\frac{1}{c^4}\left(2\left(U_{P}+V_{P}\right)^{,I}-2U^{,I}_{N}\left(U_N+V_N \right) \right)\rm{.}
 \end{eqnarray}

We can now combine these three terms together to give the full second order contribution to the deflection angle.
 \begin{eqnarray}
\label{full_2nd}
  x^{(2)I}(\vec{\theta})&&=\int^{\chi_s}_{0}d\chi\left(\chi_s-\chi \right)f^{(2)I}(\chi,\vec{\theta})\nonumber\\
 =\int^{\chi_s}_{0}d\chi\left(\chi_s-\chi \right)&&\left[
 \begin{array}{c}
\frac{1}{c^3}\left(B^{,I}_3-B^I_{,3}\right)-\frac{1}{c^4}V_{N,3}\int^{\chi}_0d\chi'\left(U^{,I}_{N}+V^{,I}_{N}\right)\\
+\frac{4}{c^4}U^{,I}_{N}\int^\chi_0d\chi'U_{N,3}+\frac{2}{c^4}V^{,i}_N\int^{\chi}_0d\chi'\left(U_{N,3}-V_{N,3}\right)\\
 +\frac{\left(U_{N}+V_{N} \right)^{,I}_{,J}}{c^4}\int^{\chi}_{0}\left(\chi-\chi' \right)
 \left[U^{,J}_N+V^{,J}_N\right]d\chi'+\frac{a}{c^4}\dot{B}^I\\
+\frac{\left(U_{N}+V_{N} \right)^{,I}_{,3}}{c^4}\int^{\chi}_{0}\left(\chi-\chi' \right)\left[U^{,3}_N-V^{,3}_N\right]d\chi'+\frac{h^{,I}_{33}}{2c^4}\\
+\frac{2}{c^4}\left(U_{P}+V_{P}\right)^{,I}-\frac{1}{c^4}h^I_{3,3}-\frac{2U^{,I}_{N}}{c^4}\left(U_N+V_N \right)
 \end{array}
 \right]\rm{.}
 \end{eqnarray}

\normalsize
It is known, see e.g. \citep{dodelson:2003}, that perturbations varying rapidly along the line of sight contribute little to the distortion
since regions of positive and negative overdensity cancel out. Thus, only modes with small $k_3$ contribute. This was considered more
quantitatively by \citep{Dodelson:2005zj} in the context of higher order lensing, where they argue that contributing modes will have
$k_3$ of order of the Hubble radius, $3000h^{-1}$Mpc. For a typical mode contributing to the deflection angle, $k_3$ will thus be at
most of order $1\%$ of $k_1$ or $k_2$. Since derivatives with respect to $x_i$ become a factor of $k_i$ in fourier space, we can neglect
the $A_{,3}$ derivatives with respect to the $A_{,I}$ derivatives for all of the potentials $A=\{U_N,V_N,B_i,U_P,V_P,h_{ij}\}$ in the
expression above, these terms will contribute negligibly to the power spectrum and can safely be ignored.
In addition, we can also ignore the last term in this expression: Considering a simple order of magnitude estimate, the scalar potential is of
order $10^{-5}$, so this term is of order $10^5$ times smaller than the first order term and we neglect it. For comparison, the Born correction 
term,$\frac{\left(U_{N}+V_{N} \right)^{,I}_{,J}}{c^4}\int^{\chi}_{0}\left(\chi-\chi' \right)\left[U^{,J}_N+V^{,J}_N\right]d\chi'$, is of order 
10\% of the leading order term \citep{Vanderveld:2011sj}.\\

 Combining the first and second order terms we arrive at the complete deflection angle
\begin{equation}
\label{bend_final}
 x^{I}(\vec{\theta})=\int^{\chi_s}_{0}d\chi\left(\chi_s-\chi \right)\left[\begin{array}{c}
\left(\frac{U^{,I}_{N}}{c^2}+\frac{V^{,I}_{N}}{c^2}\right)+\frac{2}{c^4}\left(U_{P}+V_{P}\right)^{,I}\\
+\frac{\left(U_{N}+V_{N}\right)^{,I}_{,J}}{c^4}\int^{\chi}_{0}\left(\chi-\chi' \right)\left[U^{,J}_N+V^{,J}_N\right]d\chi'\\
+\frac{B^{,I}_3}{c^3}+\frac{a\dot{B}^I}{c^4}+\frac{h^{,I}_{33}}{2c^4}
\end{array}
\right]\rm{.}
\end{equation}
This equation is the first main result of this paper. It contains all of the terms that are required for a complete description of the
weak-lensing deflection angle up to order $c^{-4}$ on non-linear scales. Since the derivation of this equation is purely geometrical,
it should hold for all metric theories of gravity. For cosmological observables to be computed, it should be supplemented by a set of
equations that relates the metric perturbations to the matter inhomogeneities for the chosen gravity theory. In section \ref{sec_matter},
we present the equations (\ref{complete_matter_eqns1})-(\ref{complete_matter_eqns8}) that relate the metric perturbations to the matter
inhomogeneities for a GR $\Lambda$CDM cosmology. 

\subsection{Comparison to previous results}
We can compare our final deflection angle (\ref{bend_final}) to the deflection angle in \citep{Dodelson:2005zj}, where the deflection angle
was calculated according to second order perturbation theory\footnote{There are several other notable works on second order perturbation theory. In particular,
\citep{bonvinfullsky} provides a comprehensive full-sky treatment. In the small-angle limit, the results in that paper agree with those of
\citep{Dodelson:2005zj}, to which we compare our results.}. The key difference is the regime of validity of the two approaches, in the sense
that the deflection angle in \citep{Dodelson:2005zj} uses perturbation theory, and thus requires all of the metric and matter inhomogeneities
to be small whereas, in our approach, the density field can be non-linear. This will be more evident in section \ref{sec_matter} when we examine 
the equations relating the metric perturbations to the matter inhomogeneities for a $\Lambda$CDM cosmology. In particular, the metric perturbations
in \citep{Dodelson:2005zj} cannot be related to matter inhomogeneities on non-linear scales in the universe, such as those in an N-body simulation.
Nonetheless, there are similarities in the two deflection angles: as is well known, the linear perturbation theory scalar potential and leading
order non-linear Newtonian potential contribute to the deflection angle in the same fashion. In addition, the only contributing ``leading-order
squared'' term in both deflection angles is the term giving rise to the Born correction and lens-lens coupling. The form of the contributions of
the vectors and tensors to the deflection angle are the same, although the second order vector and tensors are argued to be negligible in \citep{Dodelson:2005zj}
\footnote{See also \citep{clarkson14} for a more recent examination of the contribution of vectors and tensors to the weak lensing convergence
from the point of view of perturbation theory. The same result as in \citep{Dodelson:2005zj} is found.}. The same argument does not allow the vectors
and tensors to be neglected here as we are dealing with scales where structure formation is non-linear.

For clarity we note that, in \citep{Dodelson:2005zj}, the focus is on examining the second order perturbations to the metric and expanding the geodesic
equation up to second order. We are interested in performing the same calculation here, except allowing for a fully non-linear density field. Additional
perturbative corrections have been studied more recently, such as examining the luminosity distance to redshift relationship in a perturbed universe, and
thus correctly interpreting the redshift of distant sources \citep{1308.4935,1402.1933}. Typically, these results are presented in terms of the magnification,
and few papers have fully examined the effects of these additional corrections on cosmic-shear surveys. In \citep{clarkson14}, several of these corrections
are examined in terms of their effect on the convergence and it is found that they are negligible compared to the standard first order lensing term.

The key difference between this work and \citep{Vanderveld:2011sj} is the philosophy behind the metric. In our approach, there is an explicit
connection between the metric and the matter inhomogeneities in a cosmological context. This will be important for section \ref{sec_matter}.
There are differences in the details of the deflection angle too: There is an additional second order scalar term contributing to the deflection
angle in \citep{Vanderveld:2011sj}. This is the $U_N U_{N,I}$ term that appeared in (\ref{full_2nd}) but was argued to be negligible both here and
in \citep{Dodelson:2005zj}. In \citep{Vanderveld:2011sj}, this term has been kept in case the PPN parameters are large enough to make the term
observable. Furthermore, the deflection angle in this paper includes the contribution from tensor modes.

A further difference between this work and \citep{Dodelson:2005zj,Vanderveld:2011sj} is the sign of the Born correction term, although we note that
the sign of our Born correction term agrees with other works, such as \citep{Krause:2009yr}, once the difference in convention for the scalar
potentials is taken into account.

\section{Deflection angle to observable power spectra}
\label{sec_power}
Weak gravitational lensing has multiple effects on galaxy images, namely convergence ($\kappa$), rotation($\rho$) and shear
($\gamma_1$ and $\gamma_2$). These effects are described by the distortion tensor, which is the Jacobian of the deflection angle (\ref{bend_final}).
\begin{equation}
 \psi_{IJ}\equiv\frac{1}{\chi_{s}}\frac{\partial x^{I}}{\partial \theta_{J}}\equiv\left(
\begin{array}{cc}
-\kappa-\gamma_1  &\hspace{0.5cm} -\gamma_2+\rho \\
-\gamma_2-\rho  &\hspace{0.5cm}-\kappa+\gamma_1 \\
\end{array}
\right) \rm{.}
\end{equation}
 Since a statistical analysis is required to detect weak lensing effects, the observables are the power spectra of the E- and B-modes of shear.
 These are defined to be particular combinations of the power spectra of the components of shear, $\gamma_1$ and $\gamma_2$. In the flat sky,
 small angle limit where $\psi_{IJ}$ is defined as the Jacobian of the deflection angle as above, the E- and B-mode power spectra are equal
 to the power spectra of the convergence and rotation respectively. The power spectrum of the components of the distortion tensor is defined by
\begin{equation}\label{eq_pijlm}
 \langle \tilde \psi _{IJ}(\vec l) \ \ \tilde \psi ^*_{LM}(\vec {l^{'}}) \rangle=(2\pi)^2 \delta^2(\vec l -\vec {l^{'}})P^{\psi}_{IJLM}(l)\rm{,}
\end{equation}
where $\vec{l}$ is the 2-d fourier wavenumber, conjugate to the two dimensional vector $\vec{\theta}$. The convergence and rotation power
spectra are computed from the appropriate combinations of $P^{\psi}_{IJLM}(l)$, respectively
\begin{eqnarray}
P_{\kappa}&=&\frac{1}{4}\left(P^{\psi}_{1111}+P^{\psi}_{2222}+2P^{\psi}_{1122}\right)=P_{\cal E}\nonumber\\
P_{\rho}&=&\frac{1}{4}\left(P^{\psi}_{1212}+P^{\psi}_{2121}-2P^{\psi}_{1221}\right)=P_{\cal B}
\end{eqnarray}

The E- and B-mode power spectra from the scalar potentials, both the first order term and higher order Born correction and lens-lens
coupling term, have been exhaustively studied in \citep{Krause:2009yr}, where the scalar potentials in that analysis should be replaced
by the fully non-linear Newtonian ($U_N$ and $V_N$) potentials for analysing non-linear scales. The post-Friedmann scalar potentials,
$U_P$ and $V_P$, will contribute solely to the E-mode, in the same manner as the leading order scalar term, as the term involving $U_P$
and $V_P$ contributes to the components of the distortion tensor is the same fashion as the leading order scalar term. We will now
examine how the vector and tensor modes contribute to the E- and B-mode power spectra, in order to fully catalogue which of the terms
in equation (\ref{bend_final}) contribute to the E-mode and which can contribute to the B-mode. We will see that the vectors and tensors
both produce an E-mode, but, to order $c^{-4}$, only the time derivative of the vector potential generates a B-mode.

\subsection{Vectors}
Starting from equation (\ref{bend_final}) and removing the scalar and tensor terms, we have

\begin{equation}
 x^{I}(\vec{\theta})=\chi_s \theta^{s I}=\int^{\chi_s}_{0}d\chi\left(\chi_s-\chi \right)\left[
\frac{B^{,I}_3}{c^3}+\frac{1}{c^4}a\dot{B}^I\right]\rm{.}
\end{equation}

Following \citep{dodelson:2003}, we use this expression to get the distortion in terms of the vector,
 \begin{equation}
 \psi_{IJ}=\int^{\chi_{\infty}}_{0}d\chi g(\chi)\left[
\frac{B_{3,IJ}}{c^3}+\frac{a\dot{B}_{I,J}}{c^4}\right]
\rm{,}
 \end{equation}
with $g(\chi)\equiv\chi\int^{\chi_{\infty}}_\chi d\chi'\left(1-\chi/\chi' \right)W(\chi')$, where we are now dealing with a distribution
of sources, rather than a single source, $\chi_{\infty}$ is the distance to furthest source in the survey and $W(\chi')$ is the normalised
distribution of the sources.\\
Since $B_i$ is divergence-less, the vector power spectra can be defined as follows,
\begin{eqnarray}
\label{vecpowerspect}
 \langle \tilde B_i(\vec k) {\tilde B^{*}}_j(\vec {k^{'}}) \rangle&&=(2\pi)^3\delta^3(\vec k - \vec {k^{'}})P_{ij}P_{B}(k)\nonumber\\
\langle \dot{\tilde {B}}_i(\vec k) \dot{\tilde{B}}^{*}_j(\vec {k^{'}}) \rangle&&=(2\pi)^3\delta^3(\vec k - \vec {k^{'}})P_{ij}P_{\dot{B}}(k)\rm{,}
\end{eqnarray}
where $P_{ij}=\left(\delta_{ij}-\hat{k}_i\hat{k}_j\right)$. This expression arises from noticing that, for a divergenceless vector,
$k^i B_i(\vec{k})=0$. Thus, we can construct an orthonormal basis in fourier space, $\{\hat{k}^i,e^i,\bar{e}^i\}$, such that the basis
vectors can be shown to satisfy $e_ie_j+\bar{e}_i\bar{e}_j+\hat{k}_i\hat{k}_j=\delta_{ij}$ and the vector in fourier space can be
expressed in terms of the two ``polarisations'' as $\tilde{B}_i(\vec{k})=e_i B(\vec{k})+\bar{e}_i\bar{B}(\vec{k})$. Due to the assumed
isotropy of the universe, these polarisations satisfy
\begin{eqnarray}
 \langle B(\vec{k}) B^{*}(\vec{k'}) \rangle&&=\langle \bar{B}(\vec{k}) \bar{B}^{*}(\vec{k'}) \rangle=(2\pi)^3
 \delta^3(\vec{k}-\vec{k'})P_{B}(k)\nonumber\\
\langle B(\vec{k}) \bar{B}^{*}(\vec{k'}) \rangle&&=\langle \bar{B}(\vec{k}) B^{*}(\vec{k'}) \rangle=0 \rm{.}
\end{eqnarray}
Thus, 
\begin{eqnarray}
 \langle \tilde{B}_i(\vec{k}) \tilde{B}_j^{*}(\vec{k'}) \rangle&&=\left\langle \left( e_i B(\vec{k})+\bar{e}_i\bar{B}(\vec{k}) \right)
 \left( e_j B^{*}(\vec{k'})+\bar{e}_j\bar{B}^{*}(\vec{k'}) \right)\right\rangle\nonumber\\
&&=(e_ie_j+\bar{e}_i\bar{e}_j)(2\pi)^3\delta^3(\vec k - \vec {k^{'}})P_{B}(k)\nonumber\\
&&=(2\pi)^3\delta^3(\vec k - \vec {k^{'}})P_{ij}P_{B}(k) \rm{.}
\end{eqnarray}
Using the expressions (\ref{vecpowerspect}), and taking $k_3 \ll k_1,k_2$, the power spectrum (\ref{eq_pijlm}) arising from the vector potential is
\begin{equation}
P^{\psi}_{IJLM}(l)=\int^{\chi_{\infty}}_0d\chi g^2(\chi)\frac{1}{\chi^2}\frac{l_J l_M}{\chi^2}
\left(
\frac{l_I l_L}{c^6\chi^2}P_B\left(l/\chi\right)
+\frac{a^2}{c^8}\left(\delta_{IL}-\hat{l}_I\hat{l}_L\right)P_{\dot{B}}\left(l/\chi\right)
\right)
\rm{.}
\end{equation}
\normalsize
 Taking the appropriate combinations of $P^{\phi}_{IJLM}$ yields the final expressions for the E- and B-mode power spectra, $P_{\cal{E}}$ and
 $P_{\cal{B}}$ respectively, as

\begin{eqnarray}
P_{\cal{E}}(l)&=&\frac{1}{4}\int^{\chi_{\infty}}_0 d\chi g^2(\chi)\frac{l^4}{\chi^6}
 \frac{P_B\left(l/\chi\right)}{c^6}\label{vec_emode}\\
P_{\cal{B}}(l)&=&\frac{1}{4}\int^{\chi_{\infty}}_0\! \! d\chi g^2(\chi)\frac{a^2l^2}{\chi^4}
\frac{P_{\dot{B}}\left(l/\chi\right)}{c^8}
\rm{.}
\end{eqnarray}

\subsection{Tensors}
Considering just the tensor contribution to the deflection angle, equation (\ref{bend_final}) and following the same steps as for the vector,
the tensor contribution to the distortion tensor is
\begin{equation}
 \psi_{IJ}(\vec{\theta})=\int^{\chi_s}_{0}d\chi g(\chi)\left(\frac{h^{,IJ}_{33}}{2c^4} \right)
\rm{.}
\end{equation}
 We can straightforwardly see that the rotation, and therefore the B-mode will be zero, as the contribution from tensor perturbations is
 symmetric in $I$ and $J$. The power spectrum of the convergence will be non-zero and we will compute that now.
First note that, similarly to the vectors, the tensors can be split into the two polarisations in fourier space as follows, 
\begin{equation}
 h_{ij}(\vec x)=\int^{\infty}_{-\infty}d^3 k\smallskip e^{i \vec{k} \cdot \vec{x}}\left(h(\vec{k})q_{ij}+  \bar{h}(\vec{k})\bar{q}_{ij} \right)\rm{,}
\end{equation}
where $q_{ij}$ and $\bar{q}_{ij}$ can be expressed in terms of the vectors $e_i$ and $\bar{e}_i$, where again the vectors $e_i$ and
$\bar{e}_i$ form an orthonormal basis with $\hat{k}$, such that they satisfy $e_ie_j+\bar{e}_i\bar{e}_j+\hat{k}_i\hat{k}_j=\delta_{ij}$. In terms of these vectors, 
\begin{eqnarray}
 q_{ij}&=&\frac{1}{\sqrt{2}}\left(e_ie_j-\bar{e}_i\bar{e}_j \right) \nonumber\\
\bar{q}_{ij}&=&\frac{1}{\sqrt{2}}\left(e_i\bar{e}_j+\bar{e}_ie_j \right)\rm{,}
\end{eqnarray}
and the tensor power spectra are defined by
\begin{eqnarray}
 \langle h(\vec{k}) h^*(\vec{k'}) \rangle&=&\langle \bar{h}(\vec{k}) \bar{h}^*(\vec{k'}) \rangle=(2\pi)^3 \delta^3(\vec{k}-\vec{k'})P_h(k)\nonumber\\
\langle h(\vec{k}) \bar{h}^*(\vec{k'}) \rangle&=&\langle \bar{h}(\vec{k}) h^*(\vec{k'}) \rangle = 0 \rm{.}
\end{eqnarray}
When computing the contribution to the convergence, we will need to calculate $q^2_{33}+\bar{q}^2_{33}$ whilst neglecting the contribution of $k_3$.
This works out to be $1/2$, so the convergence power spectrum from the tensors is simply
\begin{equation}
P_{\cal{E}}(l)=\frac{1}{8}\int^{\chi_{\infty}}_0 d\chi g^2(\chi)\frac{l^4}{\chi^6}
 \frac{P_h\left(l/\chi\right)}{c^8}\rm{.}
\end{equation}

\section{Relating to matter content and Newtonian simulations}
\label{sec_matter}
The strength of using the post-Friedmann formalism for weak lensing is that it is possible to connect the metric perturbations with the matter
inhomogeneities in the universe, even on scales where the density contrast is greater than unity. Here, we will show how the metric perturbations
can be calculated from fully non-linear matter perturbations in a $\Lambda$CDM universe. this section will use the application of the post-Friedmann
formalism to a $\Lambda$CDM cosmology in \citep{postf}. These equations will not neccessarily apply for a dark energy or modified gravity cosmology,
so the post-Friedmann approach would have to be applied to these cosmologies in order to fully examine lensing on non-linear scales. We leave this to future work.

We will also explain which metric quantities can be fully calculated directly from $\Lambda$CDM N-body simulations, and discuss the circumstances
in which the other quantities can be extracted. This creates the opportunity for all of the contributions to the deflection angle (\ref{bend_final})
to be calculated from standard Newtonian simulations, and thus for the complete weak lensing predictions on fully non-linear scales to be calculated.

\subsection{Newtonian Regime}
Considering the leading order Einstein equations in the post-Friedmann $c^{-1}$ expansion yields the gravitational equations in the Newtonian regime \citep{postf}
\begin{eqnarray}
&&\hspace{-0.4cm}\frac{1}{c^2a^2}\nabla^2V_N=-\frac{4\pi G}{c^2}\rho_b \delta\\ &&\hspace{-0.4cm}\frac{2}{c^2a^2}\nabla^2\left(V_N-U_N \right)=0\\
&&\hspace{-0.4cm}\frac{1}{c^3}\hspace{-0.1cm}\left[\frac{2\dot{a}}{a^2}U_{N,i}+\frac{2}{a}\dot{V}_{N,i}-\frac{1}{2a^2}\nabla^2B^N_i\right]\hspace{-0.1cm}
=\hspace{-0.1cm}\frac{8\pi G\rho_b}{c^3}\left(1+\delta \right)v_i\label{vec_source}\rm{,}
\end{eqnarray}
where $\delta$ is the fully non-linear density contrast and $\vec{v}$ is the velocity of the cold dark matter. As expected, we have the Poisson
equation for $U_N$ and the equality of the $U_N$ and $V_N$ potentials. In addition, there is a vector potential that is sourced by the Newtonian
density and velocity fields. These three equations dictate how to extract $U_N$, $V_N$ and $B^N_i$ from N-body simulations. Using the curl of
equation (\ref{vec_source}), the power spectrum of the post-Friedmann vector potential was calculated from N-body simulations in
\citep{Bruni:2013mua,longerpaper} and found to be of order $10^{-5}$ times the power spectrum of the non-linear Newtonian scalar potential
over a range of scales and redshifts. This is up to two orders of magnitude larger than predicted by second order perturbation theory
\citep{Bruni:2013mua,longerpaper}, cf. \citep{0812.1349}.  Using this result in equation (\ref{vec_emode}), we can calculate the contribution
of the vector potential to the E-mode, which will clearly be negligible compared to the signal from the leading order Newtonian scalar potential.
This is the first time that the contribution of the vector potential to the cosmological weak-lensing signal has been shown to be negligible
on non-linear scales and illustrates the power of our approach.

\subsection{Gravitational equations up to order $1/c^4$}
We now consider the gravitational equations beyond the leading order Newtonian regime. In particular, $U_P$, $V_P$ and $h_{ij}$ are given by \citep{postf}
\begin{eqnarray}
\frac{1}{c^4}\nabla^2\nabla^2V_P&&=\frac{1}{c^4}\left(\frac{1}{2}\nabla^2\nabla^2V^2_N-\frac{5}{4}\nabla^2\left(V_{N,i}V^{,i}_N\right)\right.\nonumber\\
&&\left.-2\pi Ga^2 \bar{\rho}\left[\nabla^2\left(v^2\left(1+\delta \right) \right)
-3\dot{a}\left(v_i\left(1+\delta \right) \right)^{,i} \right] \right) \nonumber\\
\frac{1}{c^4}\frac{4}{3}\nabla^2\nabla^2\left(U_P-V_P \right)&&=\frac{1}{c^4}\left({\cal A}^{j,i}_{i,j}
+8\pi G a^2 \bar{\rho} {\cal S}^{j,i}_{i,j} \right)\nonumber\\
 \frac{1}{c^4}\nabla^2\nabla^2\nabla^2 h^j_i=\frac{1}{c^4}&&\left[
\begin{array}{c}
-{\cal A}^{l,kj}_{k,li}-\nabla^2{\cal A}^{l,k}_{k,l}\delta^j_i+2\nabla^2{\cal A}^{k,j}_{i,k}\\
+2\nabla^2{\cal A}^{k}_{l,ki}\delta^{lj}-2\nabla^2\nabla^2{\cal A}^{j}_{i}\\
+8\pi G a^2 \bar{\rho}\left(-{\cal S}^{l,kj}_{k,li}-\nabla^2{\cal S}^{l,k}_{k,l}\delta^j_i+2\nabla^2{\cal S}^{k,j}_{i,k}\right)\\
+8\pi G a^2 \bar{\rho}\left(+2\nabla^2{\cal S}^{k}_{l,ki}\delta^{lj}-2\nabla^2\nabla^2{\cal S}^{j}_{i}\right)
\end{array}
\right]\nonumber\\
{\cal A}^j_i&&=2V_{N,i}V^{,j}_N-\frac{2}{3}\delta^j_iV_{N,k}V^{,k}_N\nonumber\\
{\cal S}^j_i&&=\left(1+\delta \right)\left(v_iv^j-\frac{1}{3}\delta^j_iv^2 \right)\rm{.}
\end{eqnarray}
\normalsize
These quantities cannot be extracted fully from standard N-body simulations. This is because, when solving the dynamical equations at
order $c^{-4}$, the density and velocity fields will no longer quite follow the same dynamics as in the Newtonian regime. However, it
is expected that the non-linear matter dynamics in our universe is Newtonian to a good approximation and this is borne out by the small
value of the vector potential in the N-body simulations. Thus, we propose to extract the leading order contribution to these
post-Friedmann quantities by substituting the Newtonian density and velocity fields into these equations. Of course, if these
quantities turn out to be large, then this suggests that Newtonian gravity is not as good an approximation to the matter dynamics
as is currently believed and modified N-body simulations will need to be developed. Assuming this turns out to not be the case, we
propose the following system of equations for extracting all of the metric potentials that contribute to the weak-lensing deflection
angle from $\Lambda$CDM N-body simulations:
\begin{eqnarray}
&&\nabla^2V_N=-4\pi G\rho_b \delta^N \label{complete_matter_eqns1}\\
&&V_N=U_N\label{complete_matter_eqns2}\\
&&\nabla\times\nabla^2B^N_i=-16\pi Ga^2\rho_b\nabla\times\left[\left(1+\delta^N \right)v^N_i\right]\label{complete_matter_eqns3}\\
&&\nabla^2\nabla^2V_P=\left(\frac{1}{2}\nabla^2\nabla^2V^2_N-\frac{5}{4}\nabla^2\left(V_{N,i}V^{,i}_N\right)\right.\label{complete_matter_eqns4}\\
&&\left.-2\pi Ga^2 \bar{\rho}\left[\nabla^2\left(v^{N2}\left(1+\delta^N \right) \right)-3\dot{a}\left(v^N_i
\left(1+\delta^N \right) \right)^{,i} \right] \right) \nonumber\\
&&\frac{4}{3}\nabla^2\nabla^2\left(U_P-V_P \right)=\left({\cal A}^{j,i}_{i,j}+8\pi G a^2 \bar{\rho} {\cal S}^{j,i}_{i,j} \right)
\label{complete_matter_eqns5}\\
 &&\nabla^2\nabla^2\nabla^2 h^j_i=\left[
\begin{array}{c}
-{\cal A}^{l,kj}_{k,li}-\nabla^2{\cal A}^{l,k}_{k,l}\delta^j_i+2\nabla^2{\cal A}^{k,j}_{i,k}\\
+2\nabla^2{\cal A}^{k}_{l,ki}\delta^{lj}-2\nabla^2\nabla^2{\cal A}^{j}_{i}\\
+8\pi G a^2 \bar{\rho}\left(-{\cal S}^{l,kj}_{k,li}-\nabla^2{\cal S}^{l,k}_{k,l}\delta^j_i+2\nabla^2{\cal S}^{k,j}_{i,k}\right)\\
+8\pi G a^2 \bar{\rho}\left(+2\nabla^2{\cal S}^{k}_{l,ki}\delta^{lj}-2\nabla^2\nabla^2{\cal S}^{j}_{i}\right)
\end{array}
\right]\label{complete_matter_eqns6}\\
&&{\cal A}^j_i=2V_{N,i}V^{,j}_N-\frac{2}{3}\delta^j_iV_{N,k}V^{,k}_N\label{complete_matter_eqns7}\\
&&{\cal S}^j_i=\left(1+\delta^N \right)\left(v^N_iv^{Nj}-\frac{1}{3}\delta^j_iv^{N2} \right)\rm{.}\label{complete_matter_eqns8}
\end{eqnarray}
\normalsize
where $\delta^N$ and $v^N_i$ are the Newtonian density and velocity fields as found in a $\Lambda$CDM N-body simulation. This set of
equations represents the second main result of this paper and allows all of the potentials that contribute to weak-lensing from
non-linear scales to be computed from the Newtonian density and velocity fields in an N-body simulation. The equation for the
difference between $V_P$ and $U_P$ is of particular topical interest, since the difference between the two scalar potentials
has been suggested as a smoking gun for the existence of modified gravity (see e.g \citep{modgrav_2pots1}). This assertion is
only strictly correct if the quantity $V_P-U_P$ is measured from standard N-body simulations using equation (\ref{complete_matter_eqns5}) and found to be small.

\subsection{Order of magnitude upper bound for the B-mode power spectrum}
The B-mode power spectrum is not generated at leading order by the Newtonian potential, so any effects generating this will be easier to
detect than effects that generate an E-mode. As shown above, up to order $c^{-4}$, other than the second order Born correction term that
can be computed from the Newtonian potential, the only generator of the B-mode is the time derivative of the vector potential. We wish
to get an estimate of the power spectrum of the time derivative of the vector potential, in order to determine whether the B-mode signal
may be seen by future surveys. The power spectrum of the post-Friedmann vector potential was calculated in \citep{Bruni:2013mua,longerpaper}.
Ideally, the power spectrum of the time derivative of the post-Friedmann vector potential would be extracted directly from simulations,
as has been done for the scalar potential \citep{Cai}. The time derivative of the scalar potential is calculated by taking the time
derivative of the Poisson equation and substituting for $\dot{\delta}$ using the continuity equation. An equivalent approach for the
vector potential requires substituting for $\dot{\vec{v}}$ using the Euler equation, which introduces a term $U_{N,i} \delta$. Although
we are able to extract these two fields individually, we have been unable to extract the combined term from the simulations.
Investigations \citep{longerpaper} using the DTFE code \citep{dtfecode} code suggest that, in general, extracting two fields A and
B from simulations and taking the product yields a different numerical answer to directly extracting the combined field AB from the simulations.

Here, we will consider a simple analytic approximation in order to estimate the order of magnitude of $\dot{B_i}$, and hence whether
the generated B-mode may be observable. From the simulations used in \citep{Bruni:2013mua,longerpaper}, we can see that the vector potential
is of order $3\times10^{-5}$ times the power spectrum of the scalar potential, over a wide range of scales and redshifts. As this ratio only
varies slowly over time, we will use an approximate form for the time derivative of the vector potential in order to obtain an order of
magnitude estimate of the B-mode power spectrum:
\begin{equation}
\label{bdot_approx}
 P_{\dot{B}}(k)=(3\times10^{-5})P_{\dot{\Phi}}(k)\approx\left(\frac{\dot a}{a}\right)^2\left( f(a)-1\right)^2
 (~3\times10^{-5})P_{\Phi}(k)\rm{,}
\end{equation}
\normalsize
where the metric potentials are now taken to be dimensionless, so $\Phi_{\rm{now}}=\Phi_{\rm{previous}}/c^2$ and $B_{\rm{now}}
=B_{\rm{previous}}/c^3$ and we have used the linear theory prediction to relate $P_{\dot{\Phi}}(k)$ to $P_{\Phi}(k)$, with $f(a)$ being the
logarithmic growth rate of density perturbations. Taking $(f(a)-1)^2$ to be of order 1, the B-mode spectrum from $\dot{B}$ can now be expressed as
\begin{equation}
  P_{\cal{B}}(l)=\int^{\chi_{\infty}}_0\! \! d\chi g^2(\chi)\frac{l^4}{\chi^6}
P_{\Phi}\left(l/\chi\right)\left[\frac{a^2\chi^2}{4l^2c^2}\left(\frac{\dot a}{a}\right)^2(~3\times10^{-5})\right]
\rm{.}
\end{equation}
\normalsize
Considering the term in square brackets, for the purpose of computing an approximate upper bound, we will set $a=1$ and evaluate
$\dot{a}$ for $a=1$. Since $g(\chi)\rightarrow 0$ for large $\chi$, we will also set a nominal $\chi_0$ as the largest contributing scale.
Setting $\chi^2=\chi^2_0$, we can now remove the factor in square brackets from the integral to allow the B-mode to be written in terms
of the standard E-mode power spectrum,
\begin{equation}
  P_{\cal{B}}(l)\leq P^{\Phi}_{\cal{E}}(l)\frac{\chi^2_0}{4l^2c^2}\left(\frac{\dot a}{a}\right)^2(~3\times10^{-5})\rm{,}
\end{equation}
where $P^{\Phi}_{\cal{E}}(l)=\int^{\chi_{\infty}}_0 d\chi g^2(\chi)\frac{l^4}{\chi^6}
P_{\Phi}\left(l/\chi\right)$ is the standard E-mode generated by the scalar potential. For the range of scales we are considering here,
$l$ will range from 10 upwards and $\chi$ will range from ~10 to ~1000 h$^{-1}$Mpc. To calculate the upper bound, we will set
$\chi_0=1000$ h$^{-1}$Mpc and $l=10$. Then 
\begin{equation}
\label{pbdot}
  P_{\cal{B}}(l)\leq P^{\Phi}_{\cal{E}}(l)\times10^{-8}\rm{.}
\end{equation}
The final step is to estimate how much the linear theory used in equation (\ref{bdot_approx}) underestimates the fully non-linear
time derivative of the scalar potential. We can see from \citep{Cai} that this effect is scale dependent, but the linear theory
prediction for the time derivative of the scalar potential typically underestimates its amplitude by 2-3 orders of magnitude on
non-linear scales. Since the behaviour of the vector over time is similar to the behaviour of the scalar, as shown by the almost
constant rato between the two power spectra, we will assume a similar relationship between the magnitude of the linear and
non-linear amplitude of $P_{\dot{B}}$. Thus, we consider that $P_{\cal{B}}(l)$ could be up to three orders of magnitude larger
than in equation (\ref{pbdot}). This still leaves our generous upper bound as $P_{\cal{B}}(l)\approx P^{\Phi}_{\cal{E}}(l)\times10^{-5}$,
i.e. the power spectrum of the B-mode that a future survey might see has an upper limit of $10^{-5}$ times the power spectrum of the
standard E-mode spectrum from the scalar potential. This is unlikely to be found with the next generation of surveys.

We can compare this result to the known sources of the B-mode in a $\Lambda$CDM universe. In particular, the Born-correction, lens-lens coupling
and reduced shear contributions to the B-mode are calculated in \citep{Krause:2009yr} using perturbation theory. The cosmological parameters used
for our calculation of the B-modes are slightly different, however, we can partially mitigate this by comparing the ratios of each B-mode to the 
respective standard first order scalar E-mode. In our simple estimate above, the B-modes from the vector potential goes as $l^{-2}$ times the
standard E-modes, whereas the B-mode from the Born correction and lens-lens coupling in \citep{Krause:2009yr} scales similarly to the standard
E-mode. Comparing the orders of magnitude, it appears that the B-modes from the vector can only dominate over those from the Born correction and
lens-lens coupling for small $l$ ($<100$). Since our calculation is valid only for small angles (large $l$), there are likely to be few scales where
the contribution from the vector is dominant compared to that from the Born correction and lens-lens coupling. Furthermore, the signal from the
vector will always be subdominant to the signal from the reduced shear that is calculated in \citep{Krause:2009yr}. We note that this
comparison is only an estimate, as the calculation in \citep{Krause:2009yr} is performed using perturbation theory. However, this should also be
mitigated through the comparison of the ratios of the B-mode signals to the respective E-mode signals and any uncertainities are likely to be
less than the approximations used in deriving our order of magnitude upper bound.

Thus far, we have considered only the power spectra of the lensing observables. One interesting possibility that we have not considered is whether
the post-Friedmann vector potential could be observed through the bispectrum. The bispectrum of cosmic shear has attracted increasing attention recently and
there have been several measurements, e.g. \citep{bispectobs}. A full second order perturbative analysis was carried out in \citep{bonvinbispect} (again
using the full-sky formalism). In this work they showed that the second order vector contribution to the metric contributed negligibly to the E-mode of
the bispectrum in comparison to the geometric corrections (such as the Born correction and lens-lens coupling terms), which are themselves smaller than the
contribution of non-linear physics (i.e. the non-linear evolution of the scalar gravitational potential). As the ratio of the vector potential to the scalar
potential is similar on linear and non-linear scales \citep{Bruni:2013mua,longerpaper}, it is unlikely that the post-Friedmann vector potential will be
relevant for the E-mode bispectrum. No full perturbative analysis has been done for the B-mode, however given that the geometric corrections are larger
than the vector potential for the E-mode bispectrum and for the B-mode power spectrum, it seems unlikely that the vector potential will be a dominant
contribution to the B-mode bispectrum. A full investigation of the effects of the metric generated by non-linear structure formation on the
weak-lensing bispectra is beyond the scope of this paper, and we leave it to future work to address this interesting question.

\section{Conclusion}
\label{sec_conc}
Weak gravitational lensing is rapidly becoming an important tool in cosmology and promises to constrain both the parameters composing
the concordance model of cosmology and physics beyond the $\Lambda$CDM model. Weak-lensing calculations predominantly use relativistic
perturbation theory, which works well for the largest scales. However, future surveys will yield the majority of their information on
non-linear scales, requiring the use of ray tracing through N-body simulations.

In this paper we have calculated the full weak-lensing deflection angle up to order $c^{-4}$ using the post-Friedmann formalism.
Crucially, this formalism doesn't require the matter inhomogeneities to be small, so this calculation of the deflection angle can
be used on fully non-linear scales in our universe. This deflection angle, equation (\ref{bend_final}), is the first main result of
this paper and should hold for any metric theory of gravity.

The main advantage of using the post-Friedmann formalism for lensing is that the metric potentials can be related to the matter
perturbations in the universe on non-linear scales. The second main result of this paper is the set of equations
(\ref{complete_matter_eqns1})-(\ref{complete_matter_eqns8}), derived from \citep{postf,thesis}, showing how the metric potentials
contributing to the deflection angle can be computed from N-body simulations. These equations are valid for a $\Lambda$CDM cosmology.
As an example, we have used the calculation of the post-Friedmann vector potential in \citep{Bruni:2013mua,longerpaper} to show that
the vector contribution to the E-mode power spectrum is unlikely to be detected, even on fully non-linear scales.

Thus far, we have been unable to extract the time derivative of the vector potential from the simulations. However, we have calculated
a simple order of magnitude estimate of the B-mode power spectrum that this time derivative would generate, showing it to be very small.
Thus, there are no significant sources of the B-mode spectrum on non-linear scales in a $\Lambda$CDM cosmology, allowing its use as one
control of systematics as is often done in analysis of weak lensing data. We also have not yet extracted the post-Friedmann scalar
potentials $U_P$ and $V_P$ or the tensor modes from the N-body simulations, however these are higher order than the vector potential
and so cannot be significantly larger than the vector potential if using Newtonian simulations for $\Lambda$CDM cosmologies has any validity. Thus, their
contributions to the E-mode should be equally negligible. We leave it to future work to verify by direct extraction from simulations
using equations (\ref{complete_matter_eqns1})-(\ref{complete_matter_eqns8}) that these are indeed small. If this is the case then,
for a $\Lambda$CDM cosmology, the use of ray tracing tracing through N-body simulations, taking into account only the Newtonian
scalar gravitational potential, is valid on all scales. If these quantities are found to be not small then the use of Newtonian
simulations for $\Lambda$CDM cosmologies requires serious scrutiny.

In order to investigate lensing on fully non-linear scales for dark energy or modified gravity cosmologies, equivalent equations
to (\ref{complete_matter_eqns1})-(\ref{complete_matter_eqns8}) would need to be derived. It is possible that the post-Friedmann
vector potential, and/or its time derivative, are larger in modified gravity theories (such as f(R) gravity), due to both the
modified behaviour of galaxies \citep{spinf(R)} and the intrinsically relativistic nature of a scalar field. We leave it to future
work to develop the modified Einstein equations in dark energy and modified gravity cosmologies using the post-Friedmann formalism.
Once the equations have been derived, they can be applied to modified N-body simulations in order to calculate the potentials
contributing to the deflection angle (\ref{bend_final}).

{\sl Acknowledgements.} MB and DW are supported by STFC grants ST/H002774/1,\\
ST/L005573/1 and ST/K00090X/1. We thank David Bacon for comments on the manuscript.

\appendix
\section{Useful Quantities}
\label{app_quant}
For completeness, we include here the inverse metric and Christoffel symbols associated with the metric (\ref{metric2}), as well
as the first order $x^{(1)\alpha}$, $p^{(1)\alpha}$ and $f^{(1)\alpha}$ quantities.
\subsection{Inverse Metric and Christoffel Symbols}
The inverse metric is given by
\begin{eqnarray}
 g^{00}&&\hspace{-0.6cm}=-1-\frac{2U_N}{c^2}+\frac{1}{c^4}\left(2U^2_N-4U_P \right)\nonumber\\
g^{0i}&&\hspace{-0.6cm}=-\frac{aB^{i}}{c^3}\\
g^{ij}&&\hspace{-0.6cm}=a^{-2}\left[\left(1-\frac{2V_N}{c^2}+\frac{1}{c^4}\left(2V^2_N-4V_P \right) \right)\delta^{ij} -\frac{h^{ij}}{c^4}\right]\rm{.}
\end{eqnarray}
%

The only non-zero Christoffel symbols at zeroth order are
\begin{eqnarray}
 \Gamma^0_{ij}&&\hspace{-0.6cm}=\frac{a\dot{a}\delta_{ij}}{c}\nonumber \\
\Gamma^i_{0j}&&\hspace{-0.6cm}=\frac{\dot{a}\delta^i_j}{ac}\rm{.}
\end{eqnarray}

The first order Christoffel symbols are given by
\begin{eqnarray}
 \Gamma^0_{0i}&&\hspace{-0.6cm}=-\frac{U_{N,i}}{c^2}\nonumber \\
\Gamma^i_{00}&&\hspace{-0.6cm}=-\frac{U^{,i}_N}{a^2c^2}\nonumber \\
\Gamma^i_{jk}&&\hspace{-0.6cm}=\frac{\delta^i_jV_{N,k}}{c^2}+\frac{\delta^i_{k}V_{N,j}}{c^2}-\frac{\delta_{jk}V^{,i}_N}{c^2} \rm{,}
\end{eqnarray}

and the second order Christoffel symbols are given by
\begin{eqnarray}
\Gamma^0_{00}&&\hspace{-0.6cm}=-\frac{\dot{U}_N}{c^3} \nonumber \\
\Gamma^0_{0i}&&\hspace{-0.6cm}=-\frac{2U_{P,i}}{c^4}-\frac{\dot{a}B_i}{c^4} \nonumber \\
\Gamma^0_{ij}&&\hspace{-0.6cm}=\frac{aB_{i,j}+aB_{j,i}}{2c^{3}}+\frac{2a\dot{a}\delta_{ij}U_N}{c^{3}}+\frac{2a\dot{a}
\delta_{ij}V_N}{c^{3}}+\frac{a^2\delta_{ij}\dot{V}_N}{c^{3}} \nonumber \\
\Gamma^i_{00}&&\hspace{-0.6cm}= -\frac{\dot{a}B^i}{a^2c^4}-\frac{\dot{B}^i}{ac^4}-\frac{2U^{,i}_P}{a^2c^4}
+\frac{2V_NU^{,i}_N}{a^2c^4}+\frac{2U_NU^{,i}_N}{a^2c^4}\nonumber \\
\Gamma^i_{0j}&&\hspace{-0.6cm}=\frac{B^{,i}_j-B^i_{,j}}{2ac^3}+\frac{\delta^i_j\dot{V}_N}{c^3}\\
\Gamma^i_{jk}&&\hspace{-0.6cm}=\frac{\dot{a}\delta_{jk}B^i}{c^4}\hspace{-0.06cm}+\hspace{-0.06cm}\frac{h^i_{j,k}
+h^i_{k,j}-h^{,i}_{jk}}{2c^4}\hspace{-0.06cm}+\hspace{-0.06cm}\frac{2\delta^i_{j}V_{P,k}}{c^4}\hspace{-0.06cm}
+\hspace{-0.06cm}\frac{2\delta^i_{k}V_{P,j}}{c^4}\hspace{-0.06cm}-\hspace{-0.06cm}\frac{2\delta_{jk}V^{,i}_{P}}{c^4} \rm{.}\nonumber
\end{eqnarray}

\subsection{\boldmath First order $x^{(1)\alpha}$, $p^{(1)\alpha}$ and $f^{(1)\alpha}$}
The first order $f^{(1)\alpha}$ functions are given by
\begin{eqnarray}
 f^{(1)0}(\chi,\vec{\theta})&=&-\frac{2aU_{N,3}}{c^2}\nonumber\\
 f^{(1)I}(\chi,\vec{\theta})&=&\frac{U^{,I}_{N}}{c^2}+\frac{V^{,I}_{N}}{c^2}\nonumber\\
f^{(1)3}(\chi,\vec{\theta})&=&\frac{U^{,3}_{N}}{c^2}-\frac{V^{,3}_{N}}{c^2}\rm{.'}
\end{eqnarray}

In addition, the first order deflections and p-vectors are defined as
\begin{eqnarray}
 x^{(1)\alpha}(\vec{\theta})&=&\int^{\chi_s}_{0}d\chi\left(\chi_s-\chi \right)f^{(1)\alpha}(\chi,\vec{\theta})\nonumber \\
p^{(1)0}(\chi)&=&p^{(1)0}(\chi=0)-\int^{\chi}_{0}f^{(1)0}\left(\chi'\right)d\chi'\nonumber \\
p^{(1)i}(\chi)&=&-\int^{\chi}_{0}f^{(1)i}\left(\chi'\right)d\chi' \rm{.}
\end{eqnarray}

So, in our case, the first order $p^{(1)\alpha}$ and $x^{(1)\alpha}$ vectors are given by
\begin{eqnarray}
 p^{(1)0}(\chi)&=&\int^{\chi}_0d\chi'\frac{2aU_{N,3}}{c^2}\nonumber\\
 p^{(1)I}(\chi)&=&-\int^{\chi}_0d\chi'\left(\frac{U^{,I}_{N}}{c^2}+\frac{V^{,I}_{N}}{c^2}\right)\nonumber\\
p^{(1)3}(\chi)&=&-\int^{\chi}_0d\chi'\left(\frac{U^{,3}_{N}}{c^2}-\frac{V^{,3}_{N}}{c^2}\right)\nonumber\\
x^{(1)0}(\chi)&=&-\int^{\chi}_0d\chi'\left(\chi-\chi'\right)\frac{2aU_{N,3}}{c^2}\nonumber\\
 x^{(1)I}(\chi)&=&\int^{\chi}_0d\chi'\left(\chi-\chi'\right)\left(\frac{U^{,I}_{N}}{c^2}+\frac{V^{,I}_{N}}{c^2}\right)\nonumber\\
x^{(1)3}(\chi)&=&\int^{\chi}_0d\chi'\left(\chi-\chi'\right)\left(\frac{U^{,3}_{N}}{c^2}-\frac{V^{,3}_{N}}{c^2}\right)\rm{.}
\end{eqnarray}

\bibliographystyle{plain}
\bibliography{postf_lensing}

\end{document}